# Possible origin of extremely large magnetoresistance in the topological insulator $CaBi_2$ single crystal

Yuzhe Ma[1,2], Yulong Wang[1,2], Gang Wang[1,3,4,a)]

[1] *Beijing National Laboratory for Condensed Matter Physics, Institute of Physics, Chinese Academy of Sciences - Beijing 100190, China*

[2] *University of Chinese Academy of Sciences - Beijing 100049, China*

[3] *School of Physical Sciences, University of Chinese Academy of Sciences - Beijing 100049, China*

[4] *Songshan Lake Materials Laboratory - Dongguan, Guangdong 523808, China*

**ABSTRACT**

$CaBi_2$ has been experimentally found to be a superconductor with a transition temperature of 2 K and identified as a topological insulator via spin- and angle-resolved photoemission spectroscopy, which makes it a possible platform to study the interplay between superconductivity and topology. But the detailed transport properties for $CaBi_2$ single crystal remain unexplored in experiments. Here, we systematically studied the magneto-transport properties of $CaBi_2$ single crystal grown by a flux method. $CaBi_2$ shows a magnetic-field-induced upturn behavior with a plateau in resistivity at low temperature. An extremely large and non-saturating magnetoresistance up to ~15000% at 3 K and 12 T was achieved. The possible reason for the magnetic field and temperature dependence of resistivity and extremely large magnetoresistance at low temperature was discussed by adopting the Kohler's scaling law, which can be understood by the compensation effect confirmed by the Hall Effect measurement.

[a)] E-mail: gangwang@iphy.ac.cn

The unique electronic band structure of materials is often reflected in their unusual electronic transport properties. In view of it, the discovery of extremely large magnetoresistance (XMR) accompanying a magnetic-field-induced upturn in resistivity at low temperature in nonmagnetic materials, such as α-As, transition metal dipnictides $MPn_2$ (M = Nb, Ta; Pn = P, As, and Sb), and rock salt structured LnX (Ln = La, Nd, Y, and Pr; X = Bi, Sb), has stimulated tremendous interest.[1-8] XMR means that the magnetoresistance (MR) of a material reaches a high point at low temperature, as high as $10^3\%$-$10^8\%$, without saturation under rather high magnetic field.[9] Among the possible underlying mechanisms for XMR, two ones have received much more attention. The topological protection mechanism claims that the back-scattering of electrons is forbidden by the topological states in zero field and such forbiddance is not maintained under a magnetic field giving rise to XMR, like in Dirac semimetals $Na_3Bi$,[10] $Cd_3As_2$,[11] $ZrTe_5$,[12] Weyl semimetals TaAs, NbAs, NbP,[13-16] and topological nodal line semimetals ZrSiX (X = S, Se, and Te),[17-20] $PtSn_4$, $PdSn_4$.[21-23] Another is based on the compensation between electron and hole carriers, proposed first to explain the origin of XMR in $WTe_2$.[24,25] A nearly perfect compensation between electrons and holes has been found in LaSb, YSb and so on.[5,26,27] Other mechanisms, such as magnetic-field-induced metal-to-insulator transition, open-orbit Fermi surface topology, magnetic field or temperature-induced change in the Fermi surface and so on[2,3,28] have also been proposed.

However, only a few topological insulators (TIs) were found to exhibit XMR (e.g., $TaSe_3$[29]). The bismuth-based intermetallic compound $CaBi_2$ has been predicted to be a TI, and the spin-split topological surface states intersecting the Fermi energy has been experimentally observed by spin- and angle-resolved photoemission spectroscopy.[30] What's more, it has been identified as a superconductor with a superconducting transition temperature of 2 K,[31] which provides clues to the search for topological superconductivity in bismuth-based compounds. Therefore it is desirable to further study the material. The transport properties for $CaBi_2$ single crystal have not been characterized in experiments up to now, which prevents study of the behavior of electrons near the Fermi surface under magnetic field.

In this work, we have successfully grown $CaBi_2$ single crystal by a flux method and reported its transport properties in detail. A magnetic-field-induced upturn in resistivity was observed at low temperature and under high magnetic field parallel to the $b$ axis. The XMR up to ~15000% was achieved at 3 K and 12 T. Both the XMR and magnetic-field-induced upturn in resistivity were analyzed from the viewpoints of Kohler's scaling law. The electron and hole carriers determined from Hall measurement are almost perfectly compensated, which can explain the emergence of XMR to some extent. In addition, the gourd-shaped out-plane angular magnetoresistance (AMR) indicates that the Fermi surface has anisotropy and a two-fold symmetry. Our results show that $CaBi_2$ should be a good platform to investigate the relationship among XMR, topology, and superconductivity.

$CaBi_2$ single crystals were grown by the flux method using Bi as flux (See the supplementary material for details). Plate-like crystals up to 4 × 2 × 1.2 $mm^3$ were obtained, as displayed in the inset of Fig. 1(b). The crystals are air-sensitive and stored in an argon-filled glove box. Powder X-ray diffraction (PXRD) measurement was performed on the PANalytical X'Pert PRO diffractometer with Cu $K\alpha$ radiation ($\lambda$ = 1.54178 Å) at 273 K. The chemical stoichiometry and homogeneity of the crystals were determined by using a scanning electron microscope (SEM, Hitachi S-4800) equipped with energy-dispersive X-ray spectroscopy (EDX). The magnetization and electrical transport properties (i.e., resistivity, MR, and Hall Effect) were measured on a physical property measurement system (PPMS, Quantum Design). The angle-dependent out-plane transport measurements were carried out on PPMS by using a rotation option. Samples for all measurements were cleaved from grown thicker crystals using a razor blade in the argon-filled glove box.

$CaBi_2$ crystallizes in a $ZrSi_2$-type structure, having the nonsymmorphic orthorhombic space group $Cmcm$ (No. 63), as illustrated in Fig. 1(a). It possesses a quasi-two-dimensional character, which consists of the square planar lattice of Bi1 and corrugated Ca-Bi2 layer located in the $ac$ plane and stacked along the crystallographic $b$ axis. The Bi1 layer is sandwiched between two corrugated Ca-Bi2 layers. The X-ray diffraction pattern collected on the cleaved surface of $CaBi_2$ single crystal at room temperature is shown in Fig. 1(b). All peaks can be indexed by the (0$k$0) diffractions, where k is even, indicating that the b axis is perpendicular to the plate surface. The lattice parameters determined from the PXRD data are $a$ = 4.721(2) Å, $b$ = 17.095(1) Å, and $c$ = 4.619(2) Å, agreed with the previously reported values.[31] In addition, extra diffraction peak from second phase Bi (represented by the asterisk symbol) is observed (Fig. S1), which is probably ascribed to the decomposition of $CaBi_2$ in the grinding. The EDX spectrum and elemental mapping suggest that the atomic ratio of Ca and Bi in $CaBi_2$ single crystal is close to 1:2 and each element is uniformly distributed throughout the crystal, as shown in Fig. S2. More EDX results can be found in Table SI. Figure 1(c) shows the temperature-dependent magnetic susceptibility in the zero-field-cooling (ZFC) and field-cooling (FC) modes under a magnetic field of 10 Oe perpendicular to the b axis. The superconducting transition temperature determined from the susceptibility curve is about 2 K, close to the previously reported value.[31] The superconducting shielding volume fraction (SVF) is estimated to be about 125% at 1.7 K without considering the demagnetization factor.

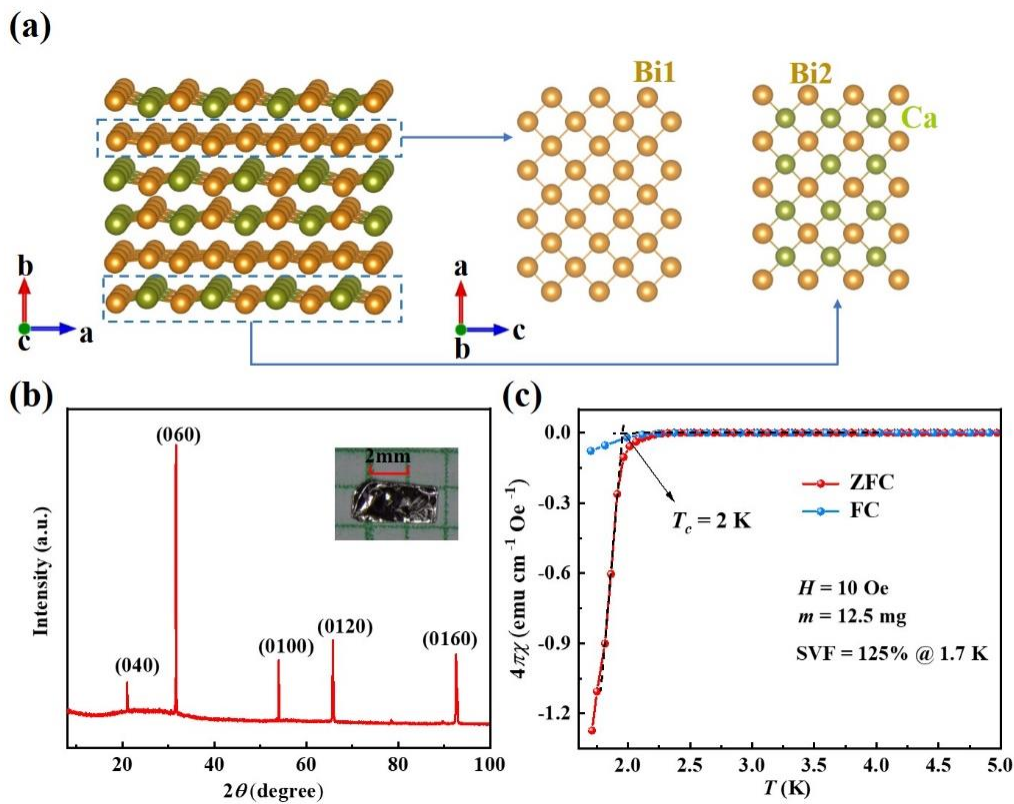

**FIG. 1.** (a) The schematic crystal structure of CaBi$_2$ viewed along the $c$ axis and the Bi1 sheet, Ca-Bi2 sheet viewed along the $b$ axis. (b) The X-ray diffraction pattern of CaBi$_2$ crystal showing (0k0) ($k$ = even) diffraction peaks. The upper inset shows the optical image of CaBi$_2$ single crystal. (c) The ZFC and FC magnetic susceptibility for CaBi$_2$ single crystal under the magnetic field of 10 Oe perpendicular to the $b$ axis.

Figure 2(a) shows the zero-field resistivity of CaBi$_2$ single crystal in the temperature range of 3 K-300 K. The resistivity shows metallic-like behavior, which decreases monotonically with the decreasing temperature from $\rho_{300\,K}$ = 47.336 μΩ cm to $\rho_{3\,K}$ = 0.219 μΩ cm. The residual resistivity ratio (RRR = $\rho_{300\,K}/\rho_{3\,K}$ = 216) suggests the high quality of CaBi$_2$ single crystal, which is comparable to that reported in ZrSiS (RRR = 288).[19] The resistivity can be described by the Bloch-Grüneisen (BG) model (equation (1)) in the temperature range of 40 K-300 K, as shown by the black solid line in Fig. 2(a), indicating that the electron-phonon scattering dominates the transport.

$$\rho(T) = \rho_0 + \left(\frac{T}{\theta_D}\right)^5 C \int_0^{\frac{\theta_D}{T}} \frac{z^5}{(e^z-1)(1-e^{-z})} dz \tag{1}$$

where $\rho_0$ is the residual resistivity at 3 K, $C$ the electron-phonon interaction constant, and $\theta_D$ the Debye temperature. The fitted parameters are $C$ = 105 uΩ cm and $\theta_D$ = 168 K. The resistivity below 40 K was fitted with the Fermi-liquid model using $\rho(T) = \rho_0 + AT^m$, where $m$ is a constant, reflecting the dominant scattering mechanism ($m$ = 2 or 5: the pure electron-electron/phonon interaction),[32] as shown in the inset of Fig. 2(a), and $m$ was determined to be 2.21, which implies the dominance of electron-electron mechanism below 40 K. The temperature-dependent resistivity under different magnetic fields in the temperature range of 3 K-100 K is depicted in Fig. 2(b). The magnetic field is parallel to the $b$ axis, and the current perpendicular to the $b$ axis. As it is shown, the resistivity in the low temperature region tends to increase with the decreasing temperature and then saturates, and the magnetic-field-induced resistivity upturn and plateau become more and more prominent with the increase of magnetic field. In TIs, the metallic surface states protected by time reversal symmetry (TRS) coexist with the insulating bulk states. The resistivity will get saturated (resistivity plateau) with the occurrence of the metallic surface conduction.[33,34] In this work, there still exists a resistivity plateau at low temperature when the magnetic field breaks TRS, excluding the scenario of topology protection. The similar behavior in resistivity at low

temperature has been observed in many topological semimetals.[35-38]

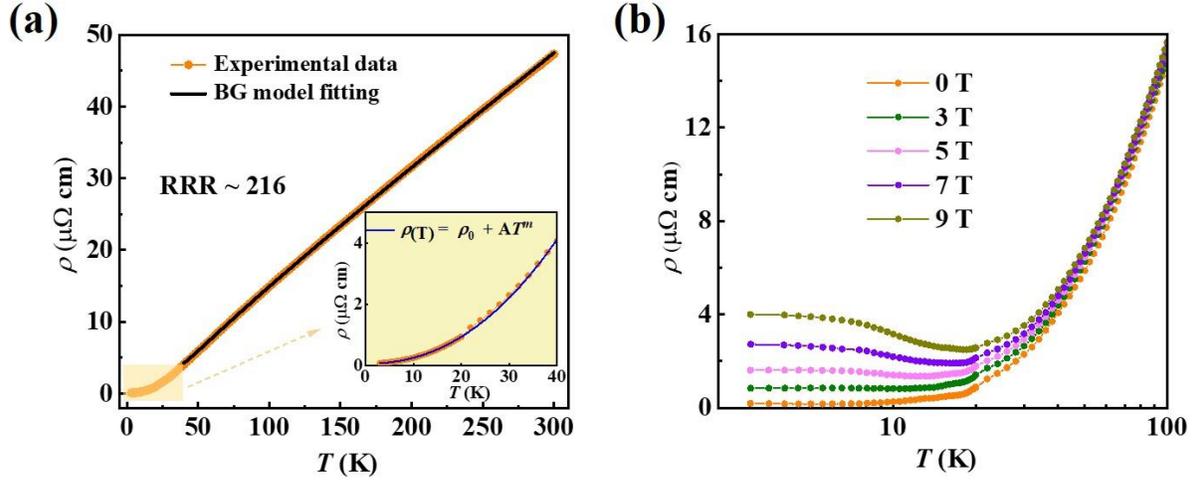

**FIG. 2.** (a) The temperature-dependent resistivity under zero magnetic field for current perpendicular to the $b$ axis. The black solid line indicates the fitting by the BG model for $T > 40$ K. The inset shows the resistivity fitted with $\rho(T) = \rho_0 + AT^m$ for 3 K $\leq T \leq$ 40 K. (b) The resistivity as a function of temperature at 3 K-100 K on a log scale under various magnetic field.

Figure 3(a) presents the magnetic-field-dependent MR at various temperatures with the magnetic field parallel to the $b$ axis and current perpendicular to the $b$ axis. MR is defined as MR = $(\rho(H) - \rho(0))/\rho(0) \times 100\%$. A MR ~15000% without any sign of saturation is observed at 3 K under 12 T. Notice that MR exhibits mainly a non-saturating $H^2$ approximate dependence, similar to many known materials with XMR, such as NbP, WTe$_2$, Cd$_3$As$_2$, TaAs and so on.[16,24,39,40] What's more, MR decreases rapidly with increasing temperature up to 50 K and reaches a value of only ~15% at 300 K under 12 T (Fig. S3). In addition, MR exhibits an oscillation at 7 K, 9 K, and 10 K, but not obvious at 3 K and 5 K. To further quantitatively analyze the oscillation behavior, a smooth background was subtracted from the $\rho(H)$ and the oscillation part as a function of $1/H$ at different temperatures are presented, as seen in Fig. 3(b). We found that the peak position of the oscillation is not linear with the $1/H$, and the amplitude cannot be fitted with the Lifshitz-Kosevich formula of the standard Shubnikov-de Hass oscillations. Further study, like angle-resolved photoemission spectroscopy under different temperature and transport measurements under high magnetic field, will be needed to determine whether it is due to the oscillation from Fermi pocket.

What's more, MR is found to be sample-dependent and closely related to the RRR of crystals. The MR of a crystal with RRR = 412 is near 3 times of that for the crystal with RRR = 170 under 9 T at 3 K, as depicted in Fig. 3(c).

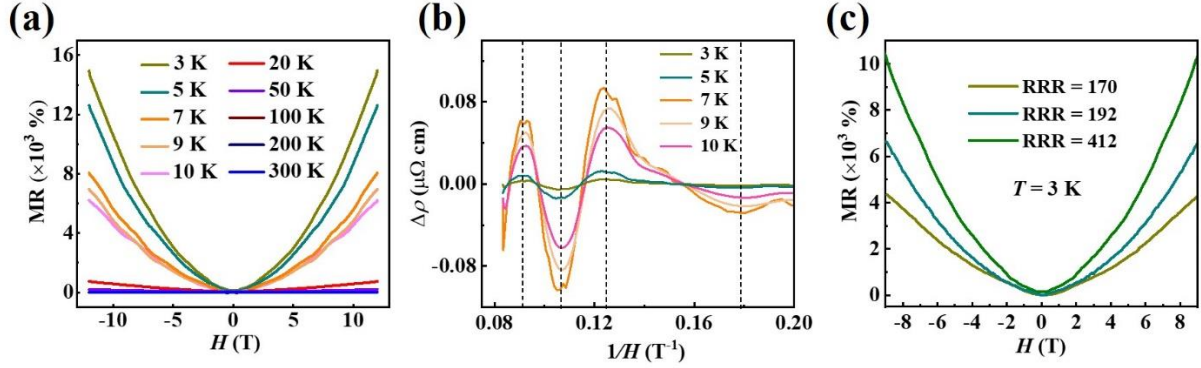

**FIG. 3.** (a) The magnetic-field-dependent MR at different temperatures. (b) The amplitude of oscillations as a function of $1/H$ at different temperatures. (c) The magnetic-field-dependent MR of $CaBi_2$ single crystals with different RRR at 3 K.

To investigate the origin of the unsaturated XMR and magnetic-field-induced upturn in resistivity, the temperature-dependent MR($T$) (MR($T$) = ($\rho(T, H)$ - $\rho(T, 0)$)/$\rho(T, 0)$ × 100%) under certain magnetic field was analyzed and shown in Fig. 4(a). The normalized MR($T$) merges into a single curve, suggesting that MR($T$) under different magnetic fields has the same temperature dependence, which excludes the existence of magnetic-field-induced metal-to-insulator transition.[41] In order to further understand the resistivity upturn, we plotted the $\rho(T, H)$ at $H$ = 0 T, 9 T and the $\Delta\rho = \rho(T, 9\text{ T}) - \rho(T, 0\text{ T})$ in Fig. 4(b) and tried to explain them with the framework of the Kohler's scaling law.[23,42] The Kohler's scaling law is used to describe the motion of electrons in magnetic field, which is governed by the following equation (2).

$$\text{MR} = \alpha(H/\rho_0)^n = (\rho(T, H) - \rho(T, 0))/\rho(T, 0) \qquad (2)$$

where $\alpha$, $n$ are constants. Here the equation (2) can be rearranged and written in the formula (3).

$$\rho(T, H) = \rho(T, 0) + \alpha H^n/\rho(T, 0)^{n-1} \qquad (3)$$

The formula clearly shows that $\rho(T, H)$ consists of two terms. The zero-field resistivity is solely determined by the vibration of temperature, and the second term caused by the magnetic field. Under a certain magnetic field, the two terms compete with each other as the temperature changes, resulting in a minimum in $\rho$-$T$ curve, shown in Fig. 2(b). As illustrated by Fig. 4(b), $\Delta\rho$ can be described well by the formula (3) with $\alpha$ = 8.32 (u$\Omega$ cm/T)$^{1.83}$ and $n$ = 1.83, which implies that the Kohler's scaling law can depict the upturn behavior and the plateau in $\rho(T, H)$ curve at low temperature. As it is known, the Kohler's scaling law is valid only for a single band or multiple bands with electron-hole compensation.[43] Wang et al. argued that this law would be violated if $\alpha$ is temperature-dependent.[24] Fig. 4(c) shows MR at 3 K, 5 K, 7 K, 9 K, and 10 K plotted against the rescaled magnetic field $H/\rho_0$. All the MR curves merge into a single curve, indicating that $\alpha$ is temperature-independent and the scattering mechanism is the same at 3 K-10 K, which further rules out the possibility of a metal-to-insulator transition.[23,44] In addition, we fitted the MR curve at 3 K with MR = $\alpha(H/\rho_0)^n$ as displayed in Fig. S4, yielding $\alpha$ = 8.31 (u$\Omega$ cm/T)$^{1.82}$ and $n$ = 1.82, which is similar to the fitted results via $\rho$-$T$ curve (Fig. 4(c)). The value of $n$ depends on the level of carrier compensation, which equals to 2 for a system with the perfect electron-hole compensation.[24,45] Thus, $n$ = 1.83 means that the electron-hole compensation should exist in CaBi$_2$ single crystal.

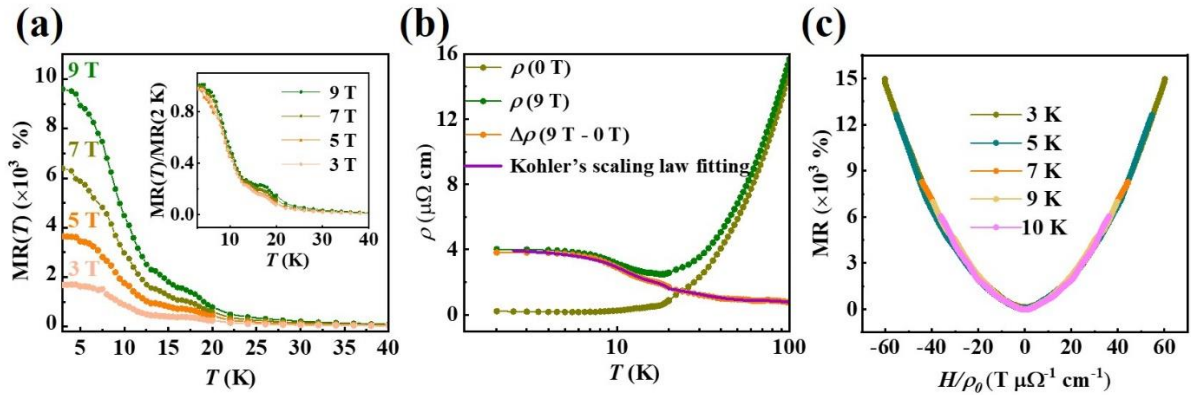

**FIG. 4.** (a) The temperature-dependent MR under various magnetic fields. The inset shows the normalized MR. (b) The temperature-dependent resistivity under 0 T and 9 T, as well as their difference $\Delta\rho$. The purple solid line indicates the fitting of $\Delta\rho$ by the Kohler's scaling law. (c) The MR as a function of $H/\rho_0$ at temperatures of 3 K, 5 K, 7 K, 9 K, and 10 K.

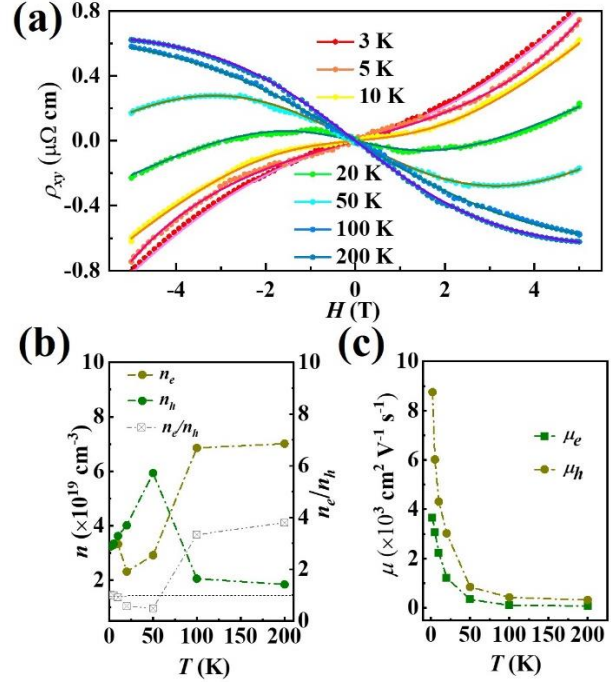

**FIG. 5.** (a) The magnetic-field-dependent $\rho_{xy}$ at different temperatures. (b) The temperature-dependent electron and hole concentration obtained from the two-band fitting. (c) The extracted carrier mobility as a function of temperature.

To determine the concentration and mobility of the carriers, the Hall resistivity ($\rho_{xy}$) at various temperature was measured, as shown in Fig. 5(a). At 3 K-10 K, $\rho_{xy}$ is found to be nonlinear and positively correlated with the magnetic field, indicating that two types of carriers are present and the hole-like carrier is dominant. On the contrary, the nonlinear $\rho_{xy}$ curve negatively correlated with the magnetic field shows that the electron-like carrier is dominant when the temperature is higher than 50 K. At 20 K and 50 K, the $\rho_{xy}$ curve changes from a negative slope at lower fields (< 3 T) to a positive one at higher fields (> 3 T), indicating that both types of carriers are involved in the Hall Effect. According to the two-carrier model, the Hall resistivity under a magnetic field can be fitted by the equation (4).

$$\rho_{xy} = \frac{1}{e} \frac{\mu_h^2 \mu_e^2 (n_h - n_e) H^3 + (\mu_h^2 n_h - \mu_e^2 n_e) H}{\mu_h^2 \mu_e^2 (n_h - n_e)^2 H^2 + (\mu_h n_h + \mu_e n_e)^2} \qquad (4)$$

Where $n_e$ and $n_h$ are the concentration of electron and hole, $u_e$ and $u_h$ the mobility of electron and hole. Subsequently, the equation can be written as follows.

$$\rho_{xy} = \frac{aH^3 + bH}{cH^2 + 1} \tag{5}$$

where *a*, *b*, and *c* can be expressed as follows.

$$a = \frac{e\mu_h^2\mu_e^2(n_h - n_e)}{\sigma_0^2} \qquad b = \frac{e(\mu_h^2 n_h - \mu_e^2 n_e)}{\sigma_0^2}$$

$$c = \frac{e^2\mu_h^2\mu_e^2(n_h - n_e)^2}{\sigma_0^2} \qquad \sigma_0 = e(\mu_h n_h + \mu_e n_e) \tag{6}$$

$\sigma_0$ is the conductivity at zero field ($\sigma_0 = 1/\rho_0$), which can be measured independently. Then, by fitting $\rho_{xy}(H)$ using equations (5) and (6), $n_e$ ($n_h$) and $u_e$ ($u_h$) at different temperatures were obtained, as displayed in Fig. 5(b) and 5(c).

At 3 K, the obtained electron and hole concentration is $3.21 \times 10^{19}$ cm$^{-3}$ and $3.24 \times 10^{19}$ cm$^{-3}$, respectively. The almost equal concentration of electron and hole implies the two types of carries are almost fully compensated with each other. As shown in Fig. 5(b), the value of $n_e/n_h$ is about 1 at low temperature and deviates from 1 with increasing temperature, indicating that the almost full compensation of electron and hole only occurs at low temperature, which is consistent with the temperature range in which XMR occurs. So the compensation of the electron and hole may be the origin of the unsaturated XMR in CaBi$_2$ single crystal. In addition, the mobility of electron and hole at 3 K is 3650 cm$^2$ V$^{-1}$ s$^{-1}$ and 8750 cm$^2$ V$^{-1}$ s$^{-1}$ respectively, which decreases with the increase of temperature, as displayed in Fig. 5(c). We also measured the AMR trying to get information of the Fermi surface, which is demonstrated to be anisotropy having a two-fold symmetry (Fig. S5).

In conclusion, we reported a careful study on the magneto-transport properties of CaBi$_2$ single crystal. A magnetic-field-induced upturn behavior and the plateau in resistivity were observed at the low-temperature region under the magnetic field parallel to the *b* axis, which can be explained in the framework of Kohler's scaling law. At 3 K and 12 T, a XMR approximated $1.5 \times 10^4$ % has been observed without any sign of saturation, which can be ascribed to the nearly perfect electron-hole compensation confirmed by the Hall measurement. In addition, the out-plane AMR shows the gourd shape with two-fold symmetry.

**ACKNOWLEDGEMENT**

Y. Z. Ma and G. Wang would like to thank Prof. X. L. Chen of Institute of Physics, Chinese Academy of Sciences, Prof. S. Jia of Peking University, and Dr. Xitong Xu of Hefei Institutes of Physical Science, Chinese Academy of Sciences for helpful discussions. This work was partially supported by the National Key Research and Development Program of China (2018YFE0202602 and 2017YFA0302902) and the National Natural Science Foundation of China (51832010).

# Supplementary material

Starting materials of Ca chips (99.5%, Alfa Aesar) and Bi lumps (99.999%, Alfa Aesar) were mixed in an alumina crucible set (Canfield Crucible Set or CCS)[1] with a molar ratio of 18:82. The alumina crucible was sealed in an argon-filled quartz tube and heated to 600 °C, after a dwell time of 10 h to get a homogeneous solution, and then slowly cooled down to 350 °C at a rate of 2 °C/h. Finally, the remaining liquid was removed by the process of a centrifuge.

**TABLE SI.** The EDX analyses for $CaBi_2$ single crystals.

| Sample No. | Point | Bi atomic ratio (%) | Ca atomic ratio (%) |
|---|---|---|---|
| 1 | ① | 66.51 | 33.49 |
|   | ② | 67.25 | 32.75 |
|   | ③ | 65.98 | 34.02 |
|   | ④ | 66.10 | 33.90 |
| 2 | ① | 65.23 | 34.77 |
|   | ② | 65.98 | 34.02 |
|   | ③ | 64.99 | 35.01 |
|   | ④ | 66.04 | 33.96 |
| 3 | ① | 66.34 | 33.66 |
|   | ② | 67.28 | 32.72 |
|   | ③ | 67.07 | 32.93 |
|   | ④ | 65.92 | 34.08 |

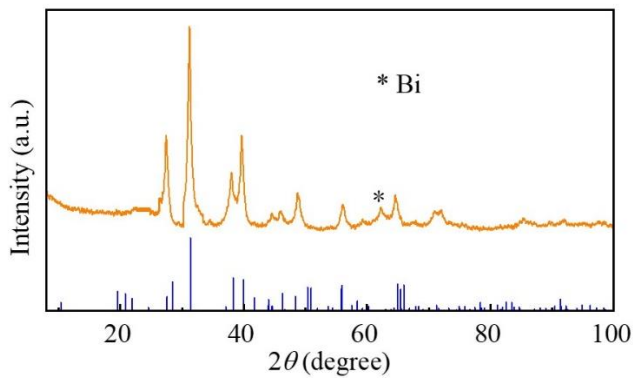

**FIG. S1.** The PXRD pattern of crushed $CaBi_2$ single crystals.

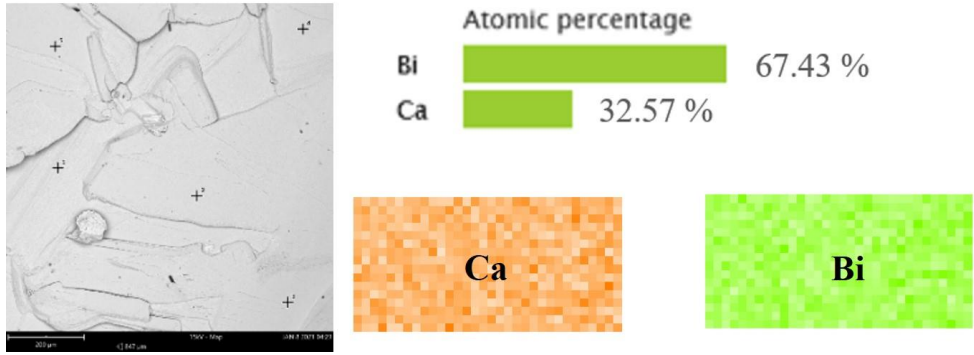

**FIG. S2.** The EDX spectrum and elemental mapping of CaBi$_2$ single crystal.

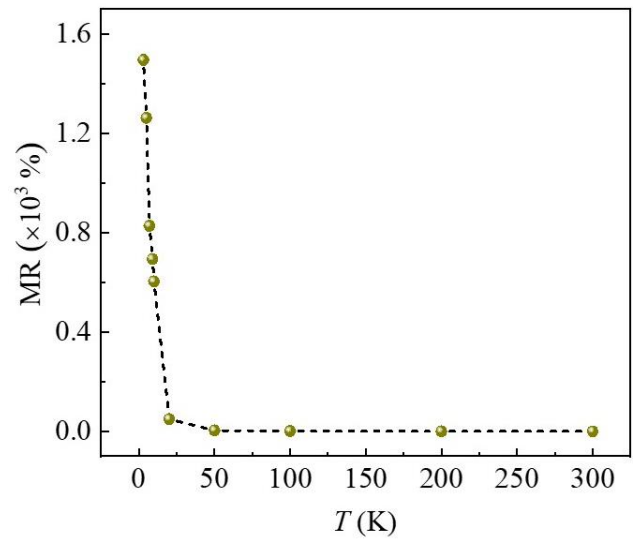

**FIG. S3.** The temperature-dependent MR at 3 K-300 K under 12 T.

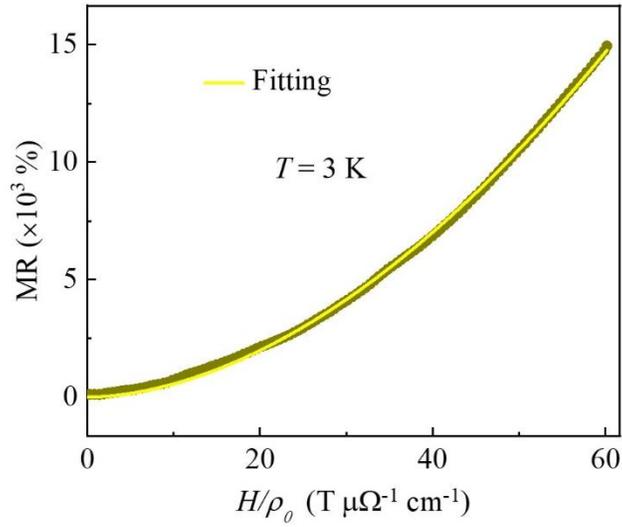

**FIG. S4.** MR versus $H/\rho_0$ at 3 K. The yellow solid line represents a fit based on MR = $\alpha(H/\rho_0)^n$, where $\alpha$, $n$ are constants.

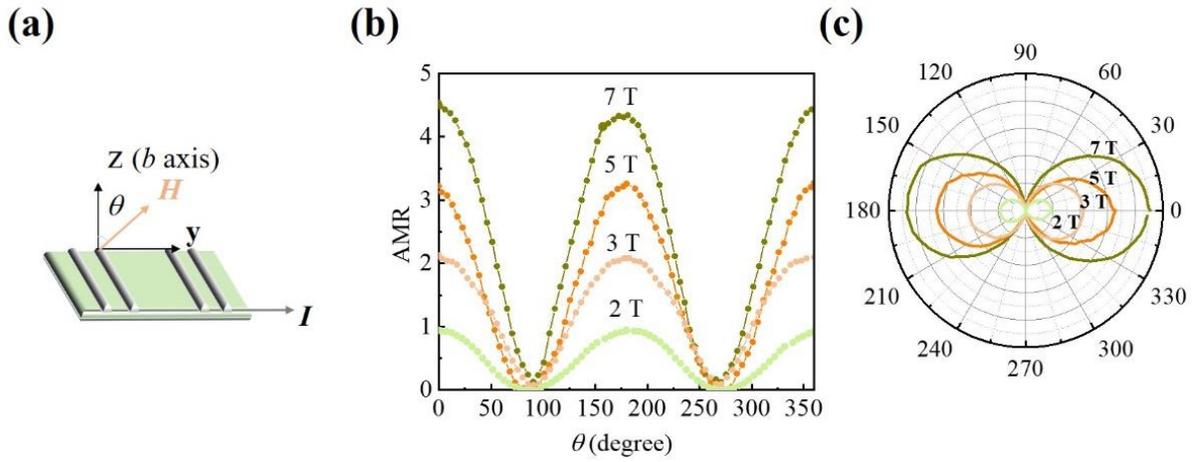

**FIG. S5** (a) The schematic of experimental configuration for AMR measurement. (b) The AMR as a function of angle $\theta$ under different magnetic fields with current perpendicular to the $b$ axis. $\theta$ is the angle between the magnetic field and the $b$ axis. (c) The AMR under different magnetic fields at 3 K presented by polar plot.

In order to gain in-depth understanding of magneto-transport properties of $CaBi_2$ single crystal, we determined the angle ($\theta$) dependence of MR by clockwise rotating the crystal in the zy plane under different magnetic fields, as displayed in Fig. S5. The experimental configuration is depicted in Fig. S5(a), where $\theta$ is the angle of the applied magnetic field with respect to the z axis.

Fig. S5(b) shows the AMR of a CaBi$_2$ single crystal under various magnetic fields ($H$ = 2 T, 3 T, 5 T, and 7 T) at 3 K. The AMR is defined by the formula.

$$\text{AMR} = [\rho_{xx}(H, \theta) - \rho_{xx}(H, 90°)]/\rho_{xx}(H, 90°)$$

It can be seen that AMR is anisotropic and reaches the maximum at $\theta$ = 0° and 180° under all magnetic fields. With increasing $\theta$, AMR decreases from $\theta$ = 0° and reaches zero at $\theta$ = 90°, which may be attributed to the anisotropy of Fermi surface.[2-4] As shown in Fig. S5(c), the polar diagram of AMR possesses a two-fold symmetry, indicating the anisotropy of Fermi surface.